\documentclass[10pt,english]{article}
\usepackage{amsmath}
\usepackage{amsfonts}
\usepackage{lmodern}

\usepackage[T1]{fontenc}
\usepackage[latin9]{inputenc}
\usepackage{amstext}
\usepackage{amssymb}
\usepackage{graphicx}
\usepackage{babel}

   \usepackage{multirow}
   \usepackage[labelfont=bf,labelsep=period,justification=raggedright]{caption}
   \makeatletter
   \renewcommand{\@biblabel}[1]{\quad#1.}
   \makeatother
\newcommand{\bv}{{\bf v}}
\newcommand{\bu}{{\bf u}}
\newcommand{\bs}{\boldsymbol{\sigma}}
\newcommand{\cI}{{\mathcal I}}

\begin{document}
\title{Stereotypical escape behavior in {\em Caenorhabditis elegans}
  allows quantification of nociceptive stimuli levels}

\author{Kawai Leung$^1$, Aylia Mohammadi$^2$, William S.\ Ryu$^{2,3,\ast}$,\\
  and   Ilya Nemenman$^{1,4,\ast}$\\
  $^1$ Department of Physics, Emory University, Atlanta, GA 30322, USA\\
  $^2$ Department of Physics, University of Toronto, Toronto, ON, M5S1A7, Canada\\
  $^3$ Donnelly Centre, University of Toronto, Toronto, ON, M5S3E1 Canada\\
  $^4$ Department of Biology, Emory University, Atlanta, GA 30322, USA\\
  $^\ast$ Email: wryu@physics.utoronto.ca, ilya.nemenman@emory.edu}
\maketitle

\begin{abstract}
  Experiments of pain with human subjects are difficult, subjective,
  and ethically constrained. Since the molecular mechanisms of pain
  transduction are reasonably conserved among different species, these
  problems are partially solved by the use of animal models.  However,
  animals cannot easily communicate to us their own pain levels. Thus
  progress depends crucially on our ability to quantitatively and
  objectively infer the perceived level of noxious stimuli from the
  behavior of animals. Here we develop a quantitative model to infer
  the perceived level of thermal nociception from the stereotyped
  nociceptive response of individual nematodes {\em Caenorhabditis
    elegans} stimulated by an IR laser. The model provides a method
  for quantification of analgesic effects of chemical stimuli or
  genetic mutations in {\em C.\ elegans}. We test the nociception of
  ibuprofen-treated worms and a TRPV (transient receptor potential)
  mutant, and we show that the perception of thermal nociception for
  the ibuprofen treated worms is lower than the wild-type. At the same
  time, our model shows that the mutant changes the worm's behavior
  beyond affecting nociception. Finally, we determine the stimulus
  level that best distinguishes the analgesic effects and the minimum
  number of worms that allow for a statistically significant
  identification of these effects.
\end{abstract}

\section*{Introduction}

Pain is a common health problem that causes over \$60 billion a year
in productivity losses in the United States
\cite{Stewart2003}. Research on pain sensing (also known as
nociception) with human subjects is difficult for a number of
reasons including ethical constraints, difficulties in quantifying a
psychophysical response, and subjectivity in self-reporting
\cite{Mogil2009}. Since the molecular mechanisms of pain transduction
are believed to be partially conserved among many different species
\cite{Tobin2002,Tobin2004, Smith2009gw}, some of
these problems are solved by using animal models in nociception
research. However, animal subjects cannot communicate their perceived
pain levels to us in an obvious fashion.  Thus progress in using
animal models depends crucially on our ability to quantitatively and
objectively infer the perceived level of noxious stimuli from animal
behavior.

Historically, studies of pain primarily used mammalian models
\cite{Barrot2012,LeBars2001}. For rodents, the tail flick test
\cite{DAmour1941}, the hot-plate test \cite{OCallaghan1975,WOOLFE1943}
and the Hargreaves' method \cite{Hargreaves1988} correlate pain
perception with the reaction latency of different body parts to
noxious stimuli. For larger animals such as canines \cite{Wegner2008}
and primates \cite{Alreja1984,Dykstra1986}, similar nociceptive assays
have also been developed. In recent years, new approaches started
incorporating facial expressions in nociception quantification
\cite{Langford2010cm, Keating:2012hq, Gleerup2015ii}. Although these mammalian models
are extensively studied, several drawbacks hinder their use. First,
ethical issues and risks arise for certain experiments. Second,
compared to lower-level organisms, mammalian subjects require more
time and resources to maintain. Therefore, much effort has been
devoted to investigations of the possibility of using invertebrate
models in nociception research \cite{Tobin2004, Im:2012gx}. In
experiments involving $Drosophila$ larva, measures such as the
response percentage of the total population
\cite{Sokabe2008,Zhong2010} and the time to response \cite{Tracey2003}
have been used to investigate changes in the ability of the animals to
sense noxious stimuli. In experiments on {\em Caenorhabditis elegans},
behavioral features such as the turning rate \cite{Tobin2002} and the
percentage of escape response \cite{Liedtke2003} have been used to
characterize nociception.

All of these models share some common problems. First, the
nociceptive assays focus on one particular coarse behavioral feature
of the subject, such as avoidance behavior, orientation, or turning
rate. Such features are selected in an ad hoc fashion, subject to a
particular design of an experiment. This makes it difficult to compare
results across different labs and experiments. Further, the behavior
may be providing additional information about the perceived pain that
is not being captured by the coarse measures.  Second, some assays
report measurements as a percentage of a population, so that these
measurements cannot be made for individuals. To overcome these
problems, an ideal assay would infer a perceived pain level of an
individual animal on a continuous scale, using comprehensive,
objective measurements of its behavioral profile.

An even more important problem is that one would like to use the
assays to calibrate the perceived pain level, and maybe even
reductions in such levels due to effects of analgesic drugs, or
mutations in the nociceptive pathways. At the same time, drugs or
mutations can affect the motor response, rather than the nociception
per se. Thus traditional pain assays mentioned above may convolve the
perceived pain reduction, if any, with behavioral changes. For
example, a mutant defective for turning behavior will register a
strong reduction in the turning rate, but it would be a mistake to
interprete this as a reduction in the pain perception. Such concerns
are very real, as is illustrated by a known fact that opiods can cause
large behavioral changes, and thus they have effects beyond analgesic
ones \cite{Roughan2000bm}. To attribute a behavioral response
difference to reduced nociception and not to motor changes, the
response must be stereotyped and reflexive, which is often the case
\cite{LeBars2001}. Further, only the response amplitude or frequency,
but not the detailed temporal structure, should change in response to
a drug or a mutation. Establishing the stability of the stereotypic
response pattern requires analysis of the {\em entire response
  behavior}, rather than of its few selected features, as is done by
most behavioral assays.

In this work, we address these issues in the context of the nematode
{\em C.~elegans}, developing it further as an animal model system for
nociception research. The worm is a great model organism for such
studies for a number of reasons. First, the behavioral dynamics of
freely moving {\em C.~elegans} is intrinsically low dimensional
\cite{Stephens2008}. This makes quantification of its behavioral
response relatively straightforward, providing an opportunity to use
the entire motile behavior as a basis for assays. Second, the worms
show a noxious response to a wide range of sensations including
certain types of chemical \cite{Culotti1978ul,Hilliard2004ig},
mechanical \cite{Kaplan1993jz,Chalfie1985vl}, and thermal
\cite{Wittenburg1999, Mohammadi:2013ku} stimuli, and such nociceptive
response is different from and is transduced independently of the
related taxis behaviors \cite{Bargmann1991, Troemel1997, Huang1994,
  Mori1995bp}.  Third, at the molecular level, many details of thermal
nociception in the worm may be similar to vertebrate animals
\cite{Tobin2004}. Fourth, there are powerful genetic and optical tools
to reveal mechanisms of nociception in {\em C.\ elegans}. Finally, the
low cost, small size, and absence of ethical constraints make the
animal amenable to large scale pharmacological screens for new
analgesics \cite{Kwok2006fl}.

We present combined experimental and modeling studies that show that
the entire temporal behavioral profile during the thermal nociceptive
response in {\em C.\ elegans} is highly stereotypical, with the
frequency of the escape response and the amplitude of the escape
velocity profile scaling with the stimulus level. By verifying the
ability of the behavioral template to capture the response following a
nociceptive stimulus, the model we develop distinguishes changes in
nociception from changes to the motor program.  When a change is
attributed to nociception, the model can infer the reduction in the
perceived thermal nociception level following pharmacological or genetic
treatments from the behavior of an individual worm. This
quantification requires only a handful of worms (about 60) to show
statistically significant nociception reduction for a common
analgesic, and its statistical power quickly improves with an
increasing number of subjects.

The paper is organized as follows. First we discuss the structure of
the dataset and the model. Then we evaluate the performance of the
model and discuss the stereotypical behavior we discovered.  Further,
we use the model to infer the nociceptive stimulus level of worms in
three different conditions: wild-type untreated, wild-type treated
with ibuprofen, and a mutant with defects in thermal nociception. We
argue that, while effects of the ibuprofen treatment can be attributed
largely to a reduced nociception, the mutant's response shows changes
to the behavior beyond nociceptive effects.  Finally, we use the
statistical model to discuss how the nociception experiments should be
designed to achieve the highest statistical power to quantify
analgesic effects.

\section*{Results}

We aim to infer the perceived worm nociception level from the temporal
dynamics of the worm response. The noxious signal is administered by thermal
stimulation using an infrared laser while the worm crawls on an agar
plate. The worm motion is captured by video microscopy and analyzed
using custom image analysis software. The worm postures are very
stereotypical, adding up to simple sinusoidal motions forwards or
backwards, and to turns \cite{Stephens2008}. Thus without much loss to
the statistical power, we characterize the entire nociceptive behavior
of the animal by a time series of its center of mass velocity (see
{\em Materials and Methods}). Our task is then to verify if such
responses are stereotypical, scaling in frequency and amplitude with
the applied laser current. If they are stereotypical and thus can be
used to characterize the perceived nociception, the next task is to infer the
applied laser current from the velocity data.

\begin{figure}[t]
\centerline{\includegraphics[width=.75\columnwidth]{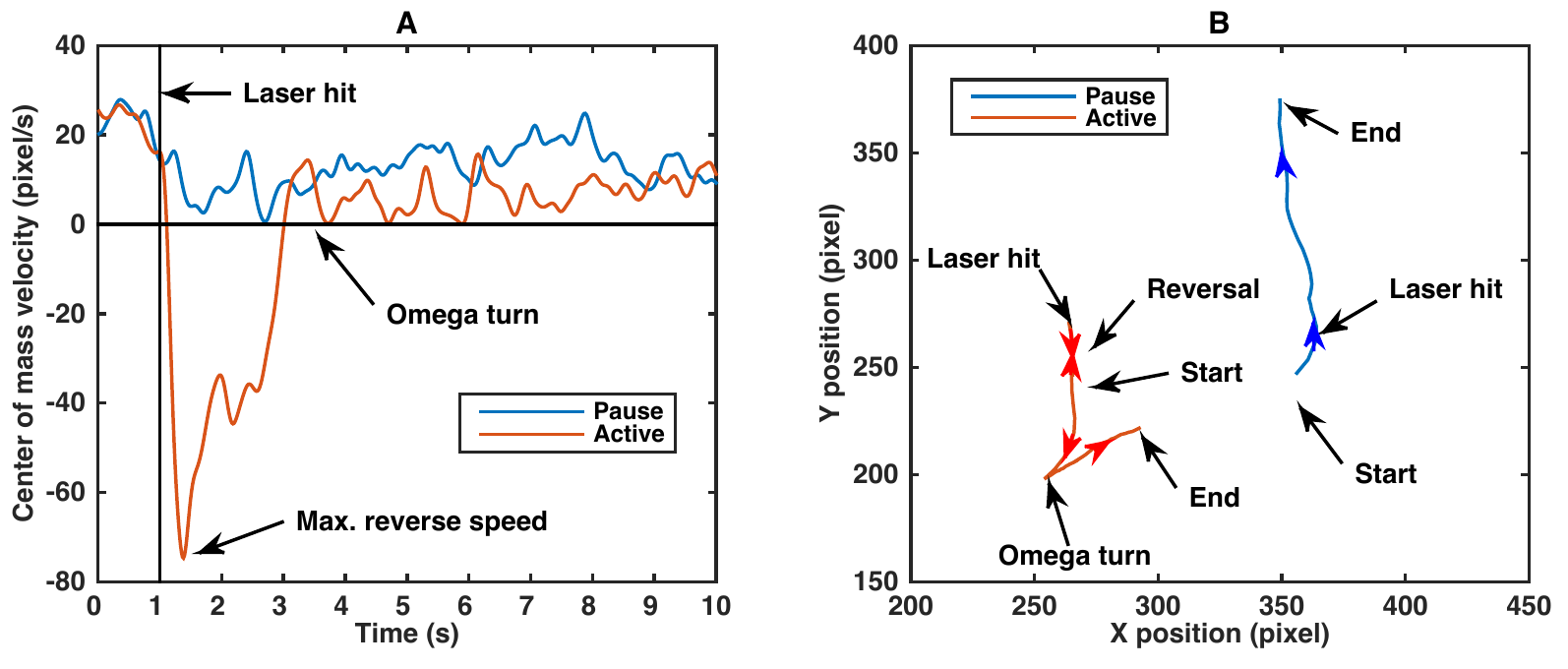}}
\caption{{\bf Typical center-of-mass thermal nociception responses in {\em
      C.~elegans}.} (A) Typical center-of-mass trajectory of two
  wild-type worms in a paused and an active state (see below for the
  detailed discussion of the two states). The infrared laser was
  directed at each worm at $t=1$ s. The paused state is characterized
  by the near zero velocity after the laser stimulus. (B) Actual
  trajectories of these two worms. Worm changes its direction of
  motion in two ways: in a ``reversal'', it stops and backtracks along
  its previous path; in an ``omega'' turn the worm's head curls back
  and crosses the tail, setting then a new direction of forward
  motion. \label{fig:CM_velocity}}
\end{figure}

For each noxious stimulus trial a random worm is selected on the plate
and its motion is sampled at 60 Hz for 15 s. An infrared laser pulse
with a randomly chosen current between 0 to 200 mA and a duration of
0.1 s is then directed to the head of the worm 1 s after the start of
the video recording. The worm's center of mass motion in a typical
trial consists of a forward motion before the stimulus, a stop and/or
backwards motion after the stimulus, then followed by an ``omega''
turn, after which the worm emerges with a forward motion in a
different direction, cf.~Fig.~\ref{fig:CM_velocity}. This is a typical
response to many noxious stimuli in {\em C.\ elegans}, and not just to
thermal noxious stimulus \cite{Mohammadi:2013ku}.  Such nociceptive
response, at least at large stimuli levels, is different from the more
commonly studied thermotactic behavior, and is mediated by different
neural and genetic pathways \cite{Wittenburg1999, Mohammadi:2013ku}.

To understand the effects of pharmacological and genetic interventions
on nociception transduction, we collected three distinct datasets. In
the first, the stimulus was applied to wild-type worms (N2). In the
second, the wild-type worms were pre-treated with an ibuprofen
solution (see {\em Materials and Methods}). In the third, we applied
the stimulus to an untreated triple mutant (ocr-2(ak47) osm-9(ky10)
IV; ocr-1(ak46)). Hereafter we refer to these datasets as ``control'',
``ibuprofen'', and ``mutant'', respectively.  We collectively refer to
ibuprofen and mutant worms as ``treated'' worms. Forward motion in all
three data sets was similar (typical forward velocity of $13\pm 9$,
$9\pm 7$, $13 \pm 10$ pixel/s respectively, where the error denoted
the standard deviation of the velocity distribution), so that there
are no drastic defects in motility.

\subsection*{Statistical model of the nociception}

\begin{figure}[t]
\includegraphics[width=1\columnwidth]{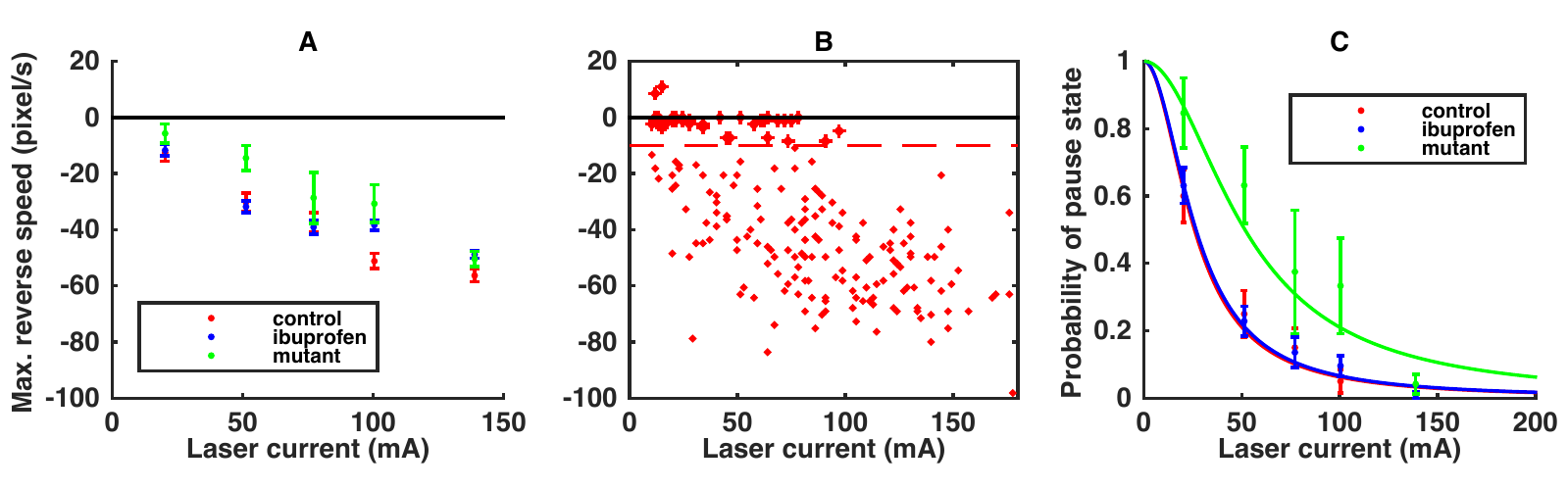}
\caption{{\bf Global characteristics of the nociceptive
    escape.} (A) The mean ($\pm$ s.\ e.\ m.) of the maximum
  reverse speed for the three worm types 
  is plotted against the stimulus laser current, partitioned into five
  bins.   Negative values correspond
  to backward motion. (B) Maximum reverse speeds for individual
  control worms. We define an active worm (dot) as having the maximum
  reverse speed of over 10 pixels/s (about 0.2 body length per second);
  otherwise the worm is paused (plus sign). (C) Probability of the paused state
  vs.\ the laser current. Dots represent the
  actual data ($\pm$ s.\ e.\ m.), and lines are the fitted
  model. The datasets are divided into equally sampled bins for each
  worm type. 
  \label{fig:P and A state demo} }
\end{figure}

We bin the laser current of the nociceptive stimulus into five
distinct levels (bins), defined to have an equal number of control
worms in each bin (40 per bin). The maximum reverse escape velocity is
indistinguishable among all three worm types for the largest stimulus
level, indicating again no gross defects to motility or noxious
response. At the same time, the velocity at smaller stimuli levels
shows substantial differences ($Z$ scores of up to 3.8) between the
control and the treated worms, especially in the vicinity of $\sim100$
mA laser stimulation, see Fig.~\ref{fig:P and A state demo}(A). Mutant
worms are especially different from the control over a wider stimulus
range.  However, as discussed above, it is unclear if such simple
observed behavioral differences are indicative of the reduction of the
perceived nociception level or of other changes to the motor response.

To address this, instead of subjectively segmenting the complex escape
behavior or choosing {\em ad hoc} metrics of the worm's movement, we
choose to infer the applied stimulus strength from a comprehensive
model of the entire worm's response velocity profile.  Due to the
considerable randomness and individual variability of responses, we
choose to model them probabilistically. Thus we are interested in
estimating $P(I|\bv)$, the probability distribution of the applied
laser current $I$ conditional on the observed velocity of the escape
response $\bv\equiv\{v(t)\}$.  Using Bayes' theorem, we write 
\begin{equation}
P(I|\bv)=\frac{P(\bv|I)P(I)}{P(\bv)}=\frac{1}{Z}P(\bv|I)P(I),
\label{eq:bayes}
\end{equation}
where $Z$ is the normalization factor, and $P(I)$ is the prior
distribution of stimuli.  When we characterize the velocity profiles
of the noxious response of {\em C.\ elegans}, we notice that the worm
can react to the stimulus in two different ways. Some worms pause
after the thermal stimulus, even at large laser currents (see
Fig.~\ref{fig:P and A state demo} (B)). These worms remains largely
immobile for a few seconds, sometimes as long as the recording
duration.  Other worms actively reverse and follow the classic escape
behavior as shown in Fig.~\ref{fig:CM_velocity}. We choose to separate
the active vs.\ the paused worms with a cutoff of 10 pixels/s, where
50 pixels is about one body length of the worm. To account for this
heterogeneity in the behavior, we introduce the state variable $s$,
which can take one of two values, $a$ or $p$, for each individual
worm. Then
\begin{equation}
P(\bv|I)=P(\bv|s=p,I) P(s=p|I)+P(\bv|s=a,I)
P(s=a|I).
\end{equation}

We model the probability of the paused state $P(p|I)$ by a sigmoid
function, cf.~Fig.~\ref{fig:P and A state demo}(C),
\begin{equation}
P(s=p|I,\cI_0)=\frac{1}{1+(I/\cI_0)^{2}},
\end{equation}
where $\cI_0$ is the {\em pause current} threshold. Then the
probability of the active state is
\begin{equation}
P(s=a|I,\cI_0)=1-P(s=p|I,\cI_0) = \frac{(I/\cI_0)^{2}}{1+(I/\cI_0)^{2}},
\end{equation}
We infer $\cI_0$ from data by maximizing
$\prod_{i=1}^{N_{\rm type}}P(s_i|I_i,\cI_0)$, where $N_{\rm type}$ is
the number of trials with worms of the analyzed type, and $I_i$ is the
actual laser current for a particular trial. Note that each of the
three data sets has its own pause current ($25.9\pm2.8$, $26.6\pm1.9$,
and $51.6\pm6.3$ mA for the control, ibuprofen, and mutant worms,
respectively). Changes in this threshold, like that for the mutant,
will result in different numbers of worms responding to the same
stimulus, which can be consistent with the changes in the nociception
level, depending on whether the response profiles themselves stay
stable. This is what we investigate next. Parenthetically, we note
that the fraction of active worms is essentially the same as the
percentage of the escape response, which has been used previously to
quantify worm nociception \cite{Liedtke2003}. Here we go further and
additionally analyze the behavioral profiles of the responding worms.

{\em C.\ elegans} locomotion consists of a series of stereotyped
postures and behavioral states
\cite{Stephens2011,Stephens2010}. Further, in other animals,
nociceptive responses are stereotyped as well
\cite{LeBars2001}. Therefore, it is natural to explore if the
nociceptive response of {\em C.\ elegans} is also stereotyped,
separately for the paused and the active states. For paused worms, the
escape velocity is small and independent of the laser current, and we
model it as a multivariate normal variable,
\begin{equation}
P(\bv|p,I)=\frac{1}{(2\pi)^{\frac{T}{2}}|\Sigma_{p}|^{\frac{1}{2}}}\exp\left[-\frac{1}{2}(\bv-\bu_{p})^{\rm
    T}\Sigma_{p}^{-1}(\bv-\bu_{p})\right],
\end{equation}
where $\bu_{p}$ is the mean velocity profile of the paused worms
measured from data, which we call the {\em paused template}
velocity. $\Sigma_{p}$ is the empirical covariance of the paused
velocity, and $T$ is the total number of effectively independent time
points in the velocity profile time series, determined using the
autocorrelation structure of the profile (see {\em Materials and
  Methods}).

We expect that, in the active state, the worm escape is laser current
dependent. Specifically, we seek to represent it by a
current-dependent rescaling of a stereotypical escape velocity,
$\bv \sim f(I)\bu_a$, where $f$ is the scaling function, and $\bu_a$
is the {\em active template} velocity.  Since various features of the
worm escapes (the maximum reverse speed, the maximum reverse
acceleration, and the time to the omega turn), scale non-linearly and
saturate with the laser current (cf.~Fig.~\ref{fig:P and A state
  demo}(A)), the rescaling, $f(I)$, must be sigmoidal. Further, some
worms have nonzero velocities even at zero laser current, so that
$f(0)$ may be nonzero. Finally, the overall scale of the template can
be absorbed in the definition of $\bu_a$. The simplest scaling
function obeying these constraints has only two parameters
\begin{equation}
f(I)\equiv f_{\cI_1,\cI_{2}}(I)=\cI_1+\frac{I}{1+I/\cI_{2}},\label{eq:nonlinerfunction}
\end{equation}
where $\cI_1$ and $\cI_{2}$ are again constants, different for the three
different worm types. With this, we write the probability of a
velocity profile given the laser current $I$ for the worm in an active
state as a multivariate normal distribution
\begin{equation}
  P(\bv|a,I)=\frac{1}{(2\pi)^{\frac{T}{2}}|\Sigma_{a}|^{\frac{1}{2}}}
  \times\exp\left[-\frac{1}{2}(\bv-f_{\cI_1,\cI_{2}}(I)\bu_{a})^{\rm
      T}\Sigma_{a}^{-1}(\bv-f_{\cI_1,\cI_{2}}(I)\bu_{a})\right],
\label{Pva}
\end{equation}
where $\Sigma_{a}$ is the covariance of the average velocity profile.
We find the constants $\cI_1$ and $\cI_{2}$, $\bu_a$ and $\Sigma_{a}$
by maximizing the likelihood of the observed data (see {\em Materials
  and Methods}).

In summary, the probability of a velocity profile given the laser
current in a certain trial is
\begin{multline}
  P(\bv|I) =
  \left\{\frac{1}{1+(I/\cI_0)^{2}}\times\frac{1}{(2\pi)^{\frac{T}{2}}|\Sigma_{p}|^{\frac{1}{2}}}
    \exp\left[-\frac{1}{2}(\bv-\bu_{p})^{T}\Sigma_{p}^{-1}(\bv-\bu_{p})\right]\right\}\\
  +\left\{\frac{(I/\cI_0)^{2}}{1+(I/\cI_0)^{2}}\times\frac{1}{(2\pi)^{\frac{T}{2}}|\Sigma_{a}|^{\frac{1}{2}}}
  \exp\left[-\frac{1}{2}(\bv-f_{\cI_1,\cI_{2}}(I)\bu_{a})^{T}\Sigma_{a}^{-1}(\bv-f_{\cI_1,\cI_{2}}(I)\bu_{a})\right]\right\}.
\label{eq:fullmodel}
\end{multline}
The overall model of the experiment, Eq.~(\ref{eq:bayes}), also
includes $P(I)$. To a large extent, this is controlled by the
experimentalist, and details are described in {\em Materials and Methods}.

\subsection*{Is the nociceptive escape stereotyped?}

\begin{figure}[t]
\centerline{\includegraphics[width=.75\textwidth]{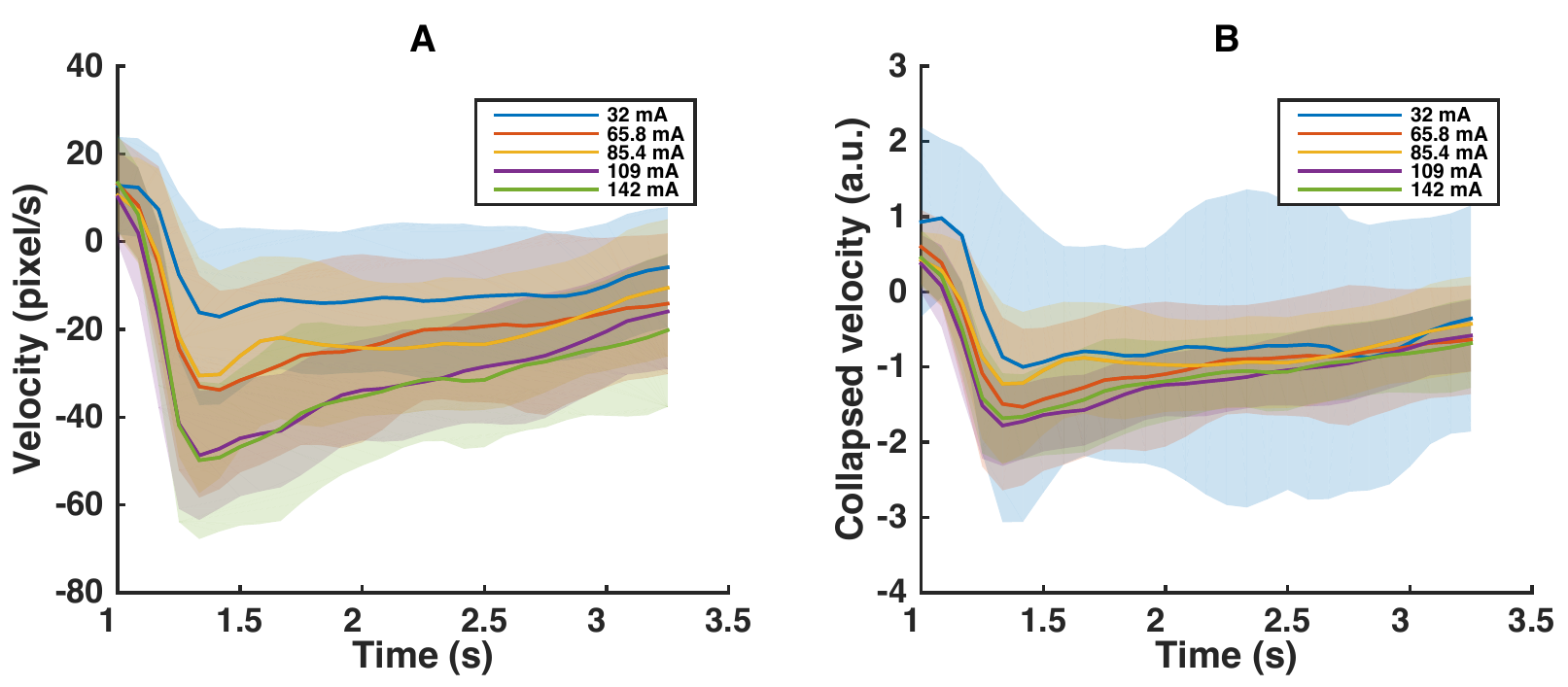}}
\caption{{\bf Collapse of the nociceptive response.} (A) The mean and
  standard deviation of velocities of active control worms, binned for
  presentation purposes into five different groups of $40$ worms each,
  based on the stimulus current. (B) The mean and standard deviation
  of velocities in the same time period rescaled by
  $f^{-1}_{\cI_1,\cI_2}(I_i)$. Rescaled mean velocities nearly
  collapse. Note that the parameters $\cI_1$ and $\cI_2$ are optimized as
  in {\em Materials and methods} to collapse individual profiles, and
  not the five mean profiles illustrated here.  \label{fig:collapse}}
\end{figure}

\begin{figure}[t]
\includegraphics[width=1\textwidth]{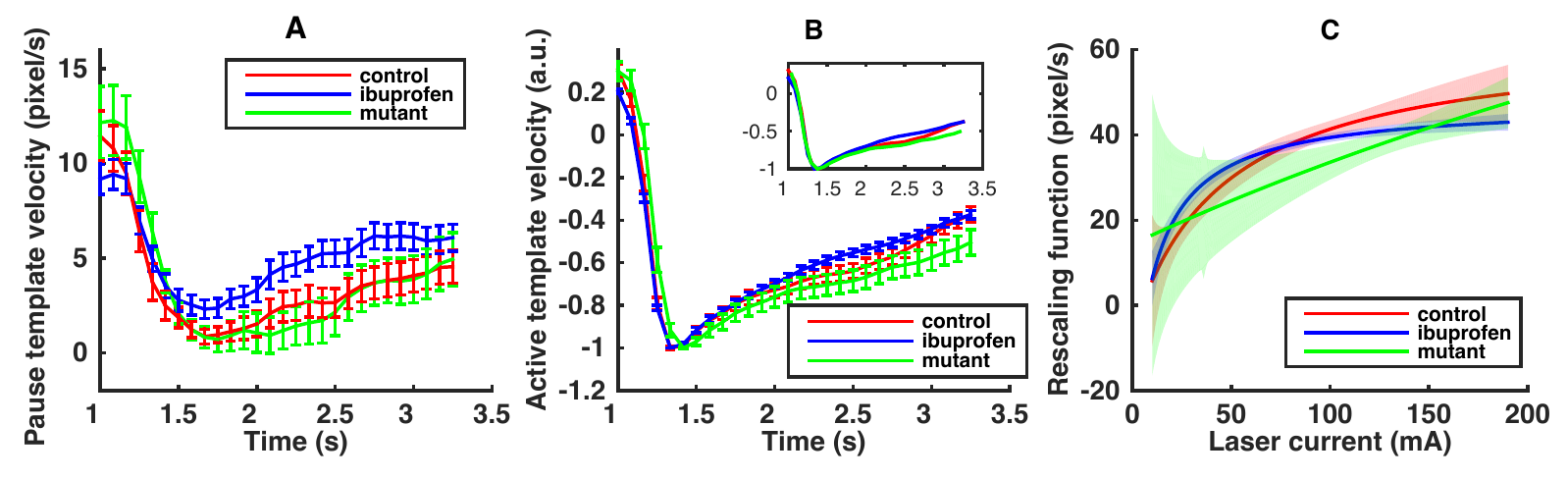}
\caption{{\bf Stereotypical active nociceptive response. } (A) Paused
  template velocities. Error bars show the standard deviation of the
  template.  Models are estimated separately for the three different
  worm types, indicated by different colors.  Time $t=1$ s corresponds
  to the moment of the stimulus application. (B) Normalized active
  template velocity.  Each template velocity is normalized by the
  absolute value of its minimum $|\min_{t} \bv_{a}|$. The error bars
  represent the model standard deviation, estimated by bootstrapping
  (see {\em Materials and Methods}). The subfigure shows the template
  velocities adjusted by the time of maximum reverse velocity,
  illustrating that the three templates nearly match. (C) The
  rescaling function $f_{\cI_1,\cI_2}(I)$ for the three different worm
  types, normalized by multiplying by the absolute value of the
  minimum of the active template velocity $|\min_{t}\bv_{a}|$. The
  optimized parameter values are $\cI_1= -4.5, -4.5, 66.5$ and
  $\cI_2=45.0, 12.0, \infty$ for the control, ibuprofen, and mutant
  worms, respectively. We deliberately do not report error bars on
  individual parameters, but rather show the one standard deviation
  confidence intervals for the entire rescaling curve as shaded
  regions on the figure. The confidence region was again estimated
  using bootstrapping.\label{fig:The-normalized-active}}
\end{figure}

The model we built assumes a stereotypical nociceptive response.  Is
this assumption justified? Velocities in the paused state are very
small (worms barely move). Thus whether the stereotypy assumption
provides a good model of the data is determined largely by the
stereotypy of the active worms.  If the active stereotypical response
template exists, then it should be possible to collapse the average
velocity profiles onto a single curve by the following transformation
\begin{equation}
\bv_{\rm collapse}=\frac{\bv_{a}}{f_{\cI_1,\cI_{2}}(I)}.
\label{eq:collapse}
\end{equation}
Figure~\ref{fig:collapse} confirms this: the means of different bins
collapse relatively compactly, providing evidence for the existence of
the stereotypy in active nociceptive responses. We illustrate the
template velocities and the non-linear scaling function $f$ inferred
from the control, ibuprofen, and mutant in
Fig.~\ref{fig:The-normalized-active}.

Note that the three active template velocity profiles in
Fig.~\ref{fig:The-normalized-active}(A) are very similar, but the
mutant template velocity profile shows a response lag of $83$ ms (one
time frame at 12 Hz) compared to the control or ibuprofen data set. In
other words, the pharmacological treatments and the mutations weakly
affect the templated response, save for possibly delaying it. This
bodes well for the assumption of the stereotypical response,
definitely for ibuprofen and, to a somewhat lesser extent, for the
mutant.

\begin{figure}
\includegraphics[width=1\textwidth]{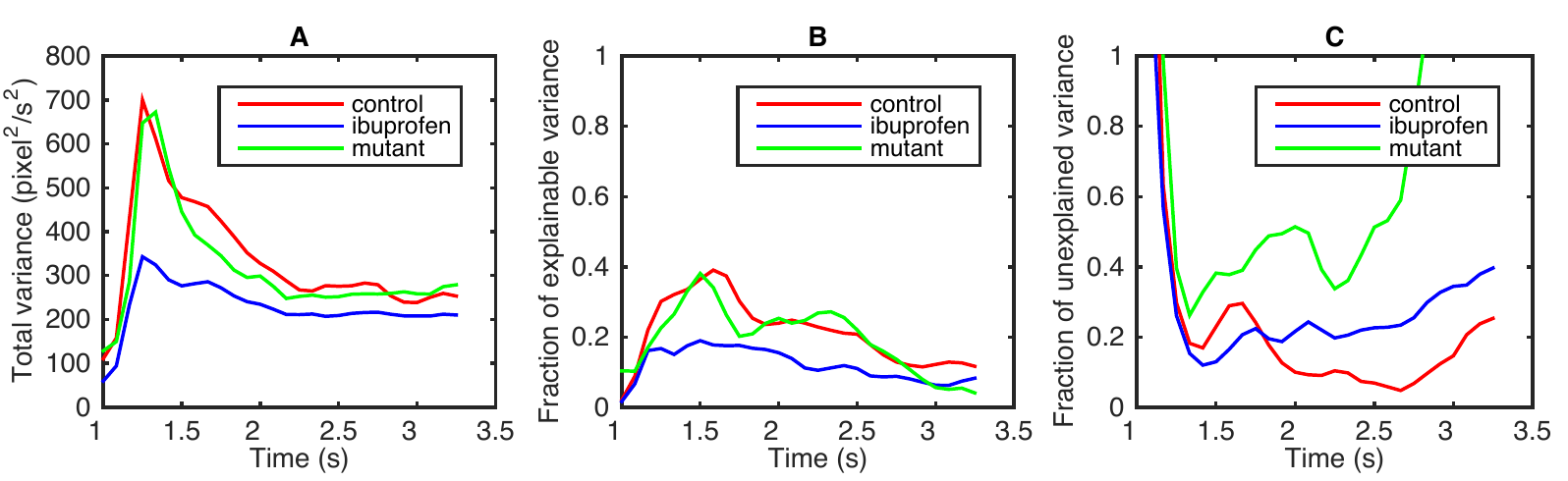}
\caption{{Variability of the active nociceptive response.} For the
  three worm types, in (A) we plot the total variance of velocity
  profiles $\bs^2_{\rm total}$. Note that these numbers depend
  strongly on the preprocessing of the data -- in particular,
  smoothing of the velocity (with the 500 ms filter, see {\em
    Materials and Methods}) decreases the total variance.  In (B) we
  show the fraction of potentially explainable variance
  $\bs_{I}^{2}/\bs_{\rm total}^{2}$. Finally, (C) shows the part of
  the explainable variance that was not explained by our statistical
  model $\bs_{\rm res}^{2}/\bs_{I}^{2}$. See the main text for the
  discussion of the differences between the three worm types on this
  plot. \label{fig:var}}
\end{figure}

While the existence of the stereotypical patterns and their similarity
across treatments is encouraging, we still need to quantify how good
the statistical models are. In the ideal case, the variance
$\bs^2_{\rm collapse}$ of the collapsed velocities,
Eq.~(\ref{eq:collapse}), calculated over individual trials, would be
zero. However, there are a number of expected sources of variance in
the velocity, such as the individual variability and the model
inaccuracies. To establish how good the stereotypical model fits are,
we need to disambiguate these contributions.  For this, we again
partition all velocity profiles into five current bins.  We then write
the total variance of all responses as
\begin{equation}
\bs_{\rm total}^{2}=\bs_{I}^{2}+\bs_{\rm ind}^{2},
\end{equation}
where $\bs_{\rm ind}^{2}$ is the variance due to individual responses
within each bin, and $\bs_{I}^{2}$ is the current-driven variance of
the mean responses across the bins.  Since the individuality of the
worms is not accounted for in our model, $\bs_{I}^{2}$ represents the
maximum potentially explainable variance in the data. The
stereotypy-based model would be nearly perfect if $\bs_{I}^2$ were to
drop to zero after the $f^{-1}$ rescaling. To explore this, in
Fig.~\ref{fig:var}, we plot the total variance of the active response
$\bs_{\rm total}^{2}$, panel (A), and the fraction of the potentially
explainable variance, $\bs^2_I/\bs^2_{\rm total}$, panel (B). The
latter varies from 20\% to 40\% of the total variance, depending on
the time post-stimulus and on the treatment. In both panels, the
mutant and the control dataset are nearly indistinguishable, while the
ibuprofen worms show a smaller variance, and a smaller fraction of the
explainable variance. This is consistent with a smaller
stimulus-driven nociceptive response for this analgesic treatment. At
the same time, these figures suggest that the decrease in the maximum
reverse speed in the mutant worm, Fig.~\ref{fig:P and A state demo}
(A), should not be attributed entirely to the reduced
nociception. Indeed, the similarity of the variance and the
explainable variance in the control and the mutant worms, which have
very different mean maximum reverse velocities, suggests existence of
an additional (explainable, non-templated) component in the
nociceptive response of the mutant, which is not present in the
control.

The explainable variance $\bs_{I}^{2}$ is further split into the
variance explained by the model, $\bs_{\rm m}^2$, and the residual
variance, $\bs^2_{\rm res}$, which the model fails to explain:
\begin{equation}
\bs_{I}^{2} = \bs_{m}^{2} + \bs_{\rm res}^{2}.
\end{equation}
In Fig.~\ref{fig:var} (C), we plot $\bs_{\rm res}^{2} / \bs_{I}^{2}$,
which is the fraction of the variance not captured by our model. Of
the explainable variance, about 80\% is captured by our model for the
control and the ibuprofen worms in the window between 1 s and 3.3 s
since the start of the trial, on average. This is a relatively large
fraction for behavioral data, and provides an additional validation
for our choice of a stereotypy-based model for representing
nociceptive escape behavior in these worms. Stability of the template
itself between these two conditions, and the stability of the fraction
of the explained variance suggest that much of the effect of ibuprofen
can be attributed to scaling of the templated response (and also the
fraction of active worms). In other words, ibuprofen decreases the
sensed nociception.

In contrast, the unexplained variance for the mutant is about twice as
large as that for the control, and approaches 100\% at $t>2.7$ s. This
again illustrates that the templated response model is not very good
for this treatment. Thus the mutations introduce changes in the
behavior that are not consistent with a simple rescaling -- mutations
affect the fine motor behavior in addition to nociception per se.

\subsection*{Using the statistical model}

One of the goals of our study is to develop methods for quantitative
assessment of the efficacy of pharmacological interventions to
decreases sensed nociception, at least in those cases where their action can
be specifically interpreted as a change in nociception transduction. We can use the developed statistical
model for this. Specifically, taking the model derived from the
control worms, we can infer the laser current from behavior of all
three different worm types. To the extent that the current inferred
for the treated worms is smaller than that for the control worms at
the same applied current, the nociception level perceived by the
treated worms is smaller.

Figure \ref{fig:PvsA prob map =000026 exp value} shows the overall
structure of the inference done with the model. In the first row, we
plot the conditional distribution of the inferred laser current given
the actual applied current $I$ for the three worm types,
$P(I_{\rm inf}|I,{\rm type})$. For presentation purposes, we again bin
the trials using $I_i$ into five bins $I_\mu$, $\mu=1,\dots,5$ as
before, and plot
\begin{equation}
  P(I_{\rm inf}|I_\mu,{\rm type})= \sum_i^{N_{\rm type}} P_{\rm
    control}(I_{\rm inf}|\bv_i)P_{\rm type}(\bv_i|I_\mu).
\end{equation}
Here $P_{\rm type}(\bv_i|I_\mu)$ is 1 if the stimulus on the $i$'th
trial for this worm type was in the $I_\mu$ bin, and zero
otherwise. Further, $P_{\rm control}(I_{\rm inf}|\bv_i)$ is given by
the full model, Eq.~(\ref{eq:fullmodel}), with the parameters inferred
for the control worm, and with the empirically observed velocities
$\bv$ in trial $i$ for each worm type. We see that there is more
probability concentrated at small $I_{\rm inf}$ for the ibuprofen and
the mutant worms, suggesting a reduction in the perceived stimulus
level.  Similarly, in the second row in Fig.~\ref{fig:PvsA prob map
  =000026 exp value}, we plot the expected value, $\bar{I}_i$, of the
distribution of the current inferred using the control model,
$P_{\rm control}(I_{\rm inf}|\bv_i)$, for each of the individual
trials in each of the three worm types. To the extent that the dots
for the ibuprofen and the mutant worms are again somewhat lower than
for the control worms, there is some reduction in the perceived
current by this measure as well.

\begin{figure}[t]
\includegraphics[width=1\textwidth]{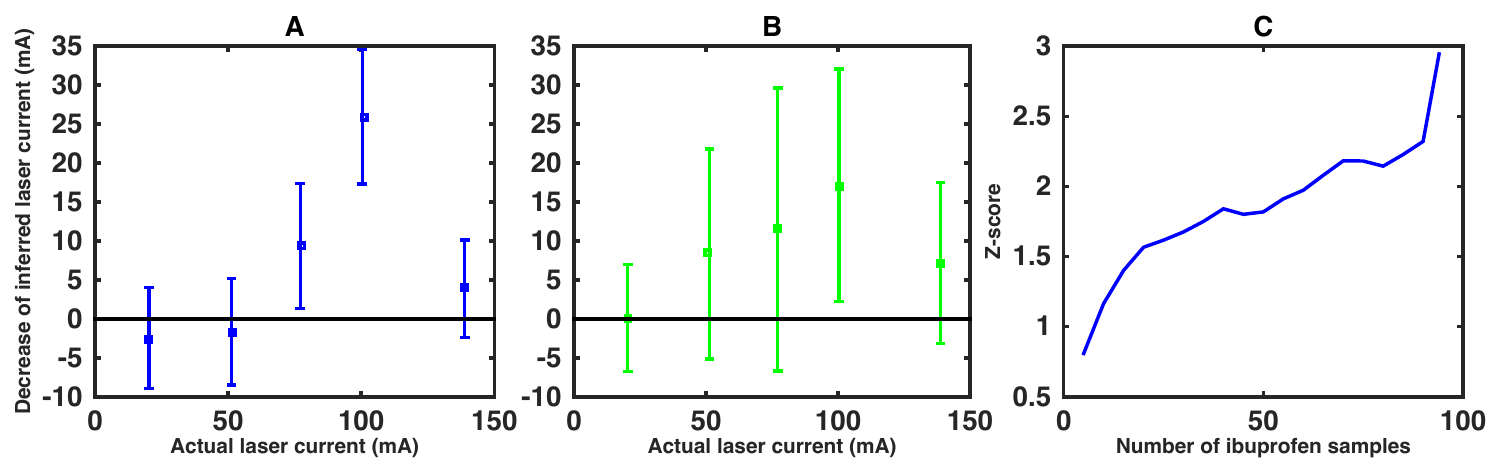}

\protect\caption{{\bf Sensed nociception decrease due to treatments.}  (A)
  Differences of the inferred laser current between the control and
  the ibuprofen datasets, $\Delta I_{\rm ibuprofen}$. Errors represent
  standard deviations, estimated with bootstrapping. (B) Similar
  differences for the mutant worm, $\Delta I_{\rm mutant}$.  Error
  bars are estimated by bootstrapping. (C) Dependence of the statistical
  significance of the nociception level reduction (measured by the $Z$ score)
  for ibuprofen at 110 mA on the number of ibuprofen trials.  
  \label{fig:Delta_Ipred}}
\end{figure}

However, these population averaged results wash out important
differences in the structure of the nociception response.  To quantify these
small effects more accurately, we now look at the perceived stimulus
changes for individual worms in the datasets. Specifically, for each
trial $i$ in the control dataset, a trial $j(i)$ with the closest
value of the applied laser current is found in the ibuprofen / mutant
dataset (the mean magnitude of the current mismatch is $<1$ mA for
both the ibuprofen and the mutant worms). We then use the control
model to calculate the expected value of the inferred current for the
$j$th trial in the ibuprofen / mutant datasets. This expectation is
subtracted from the expectation value of the inferred current for the
matched trial $i$ for the control dataset. The difference of the
expectation values, averaged over all control worms, is our measure of
the reduction in the perceived nociception level
\begin{equation}
\Delta I_{\rm type}=\frac{1}{N_{\rm control}}\sum_i^{N_{\rm control}}\left(I_{i,{\rm control}} - I_{j(i),{\rm type}} \right).
\end{equation}
We evaluate $\Delta I_{\rm type}$ for different worm types and for
controls worms binned into the five usual current bins, and show the
results in Fig.~\ref{fig:Delta_Ipred}. To estimate the error of
$\Delta I_{\rm type}$, we bootstrap the whole analysis pipeline, see
{\em Materials and Methods}. There is a statistically significant
difference in nociception perception between the ibuprofen and the control
worms.  The difference is most significant when the actual laser
current is around 100-110 mA. This coincide with our observation from
Fig.~\ref{fig:P and A state demo} (A) that the most sensitive region
of maximum reverse speed is around 100mA. Indeed, at smaller currents,
the nociception level is small, many worms pause, and the behavior
cannot be used to reliably estimate the stimulus level. At high
current, the noxious stimulus perception saturates, and all worms
behave similarly, again reducing the ability to disambiguate the
applied current level.

This analysis of ibuprofen worms achieves one of our main goals. It
proves our ability to reconstruct stimulus from the behavior, and
shows that analgesic effects of pharmacological perturbations can be
quantified from the behavior. At the same time,
$\Delta I_{\rm mutant}$ turns out to be insignificant in
Fig.~\ref{fig:Delta_Ipred} (B), even though Fig.~\ref{fig:P and A
  state demo} (A) suggest that a large statistically significant
difference exists between the mutant and the control behaviors. This
failure to detect a significant nociception reduction is because the
templated response model is not very good for the mutant worm; thus
our analysis cannot reliably assign a mutant trajectory on a given
trial to a specific nociception level. In other words, the large error bars
in Fig.~\ref{fig:Delta_Ipred} (B) serve as yet another check for self
consistency: effects of the mutations cannot be attributed just to
changes in nociception perception.

\begin{figure}[t]
\includegraphics[width=1\textwidth]{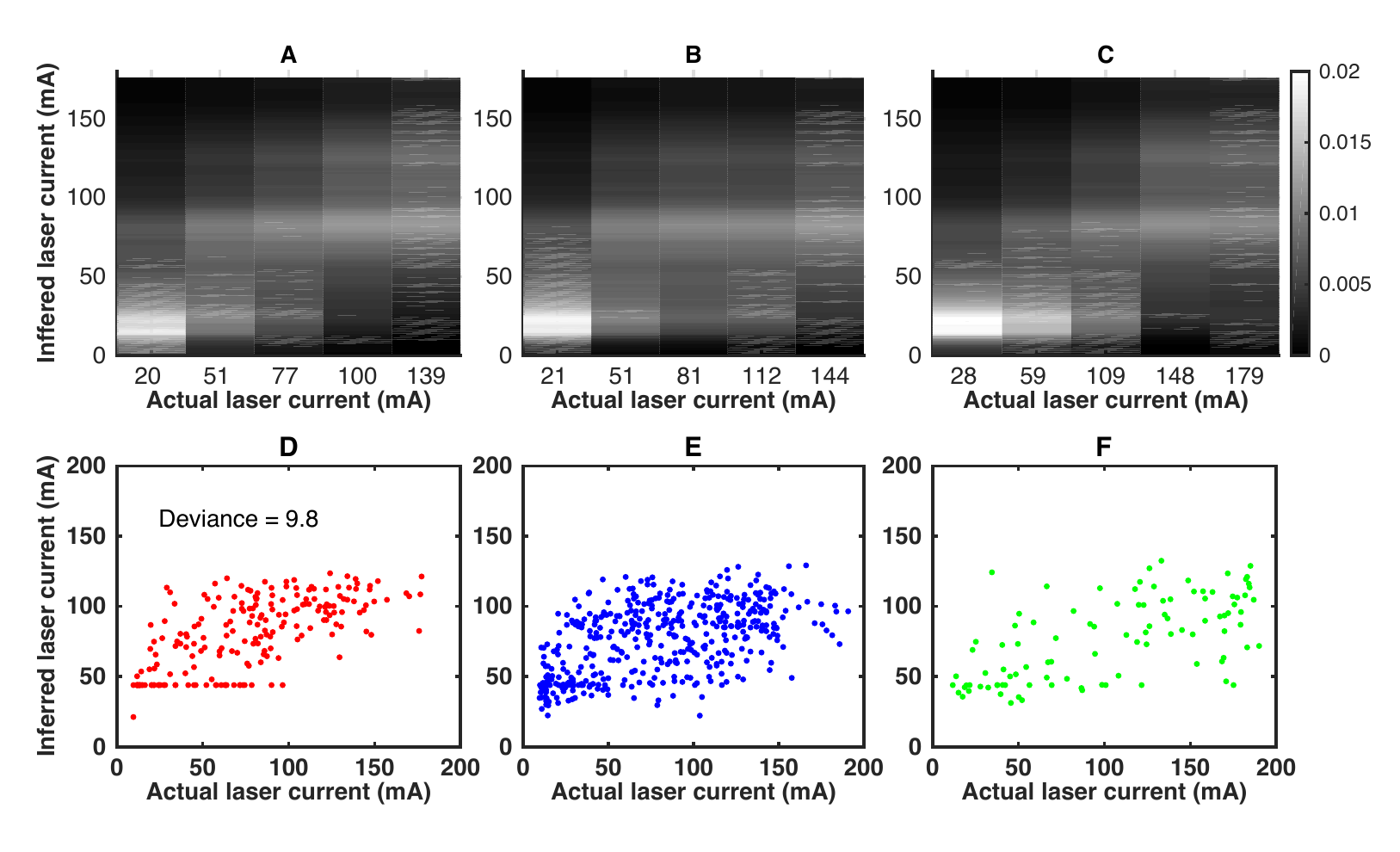}
\caption{{\bf Inferring the perceived nociception level from {\em C.~elegans}
    nociceptive behavior.} The first row shows the heat maps of the
  conditional distributions of the inferred current vs.\ the actual
  applied current, partitioned into five bins,
  $P_{\rm control}(I_{\rm inf}|I_\mu)$. Inference is done with the
  model of the control worm behavior. Panels (A), (B), (C) show the
  results for the control, ibuprofen, and mutant worms,
  respectively. The second row shows the expected inferred laser
  current for each trial, $\bar{I}_{{\rm inf},i}$ vs.\ the applied
  current $I_i$. The inference is again done using the control model,
  and panels (D), (E), (F) show the three worm types. \label{fig:PvsA
    prob map =000026 exp value}}
\end{figure}

\section*{Designing experiments: how many worms?}

We expect our analysis to be useable for screening large numbers of
chemicals for analgesic action. Since our approach targets one
individual worm at a time, we need to estimate the number of worms
needed to achieve statistical significance in such screening
experiments. For this, we fix the number of control worms, arguing
that these must be only analyzed once, and hence a relatively large
number of them can be tested. We then focus on ibuprofen, whose action
is analgesic in our experiments, and on the bin at 110 mA, where the
worms experience the most significant nociception reduction. There are
$N_{{\rm ibuprofen},110}=94$ worms in this bin. We randomly sample
with replacements $n<N_{{\rm ibuprofen},110}$ worms from among these
ibuprofen-treated worms and repeat our analysis pipeline, estimating
the $\Delta I_{\rm ibuprofen}(n)$. Resampling 1000 times (both the
ibuprofen and the control datasets), we also estimate the variance of
$\Delta I_{\rm ibuprofen}(n)$, and hence the $Z$ score as a function of
$n$, which is plotted in Fig.~\ref{fig:Delta_Ipred} (C). The plot here
is an {\em underestimate} of the true $Z$ score since resampling with
replacements removes some stimulus values from the dataset, hence
increasing the mismatch between control worms and the paired treated
worms. Even with this, $Z\approx2$ is achieved at $n\approx 60$
ibuprofen worms. In other words, in a typical screening experiment,
one would need to test $200$ or more worms to build the control model,
and then at least $\sim 60$ worms additionally for each treatment
condition.

\section*{Discussion}

Typically the goal of sensory-response experiments is to develop a
model that can predict the behavior in response to the stimuli. Here
we wanted to do this in reverse. Our goal was to build a statistical
model of the thermal stimulus from careful measurements of the
nociceptive behavior of {\em C.\ elegans}, and to use this model to
infer the changes in the perceived level of ``pain'' felt by the
organism due to perturbation in the nociception transduction
pathway. Given this model, we could then measure changes in
nociception perception due to effects of analgesics and mutations, and
use this as a basis to study the mechanism of nociception transduction in a
genetically tractable organism amenable to high-throughput screens. As
a representative data set, we choose to study the standard laboratory
{\em C.\ elegans} strain N2, N2 treated with ibuprofen, and a mutant
with defects in TRPV function.

For the model to be successful, we had to meet a number of challenges.
Since the worm could not communicate its nociception level to us directly we
had to infer this level by reading the ``body language'' of the worm's
``pain'' response. The difficulty with quantifying a behavioral response
as a measure of nociception level is that drugs or mutations can
affect locomotory behavior in addition to perturbing sensory
transduction. So in an attempt to deconvolve these effects, we used
the entire behavioral profile instead of making ad hoc measurements
and leveraged the fact that nociception responses are typically highly
stereotyped. This resulted in the discovery that the response can be
modeled with a velocity profile template that scales non-linearly in
response to an applied laser current. The success of the template in
modeling the stereotyped wild-type nociception response was confirmed
by a functional collapse of the velocity profiles across different
nociception levels. This discovery of invariance is important since it
not only allowed us to effectively correlate nociceptive behavior to
the stimulus level, but it also allowed us to determine if the
locomotory changes in our thermal nociception assay was due to changes
specifically in the sensory transduction pathway or due to other
general locomotory factors. By carefully accounting for the variation
in our data and quantifying how much of this variation is captured by
the model, we showed that the stereotypical behavior is unaffected by
ibuprofen, save for changing the amplitude of the response. Thus this
drug application likely reduced the perceived nociception level in the
worm. In contrast, a TRPV mutation changes locomotion in a way that is
not as well captured by the template model. Thus we can be objectively
critical about any nociception inference made with this strain.

The model was also useful in determining key experimental parameters
for future measurements. After verification that the model works well
with the native and ibuprofen treated stimulus-response data, we
quantified the changes in thermal nociception perception due to
ibuprofen treatment.  Our modeling and experimental assessment of
thermal nociception identified the optimal stimulus range and required
number of trials to determine statistically significant differences
between the inferred current of N2 in the untreated and treated
conditions.

In conclusion, we have built a general model that connects stereotyped
behavior to stimulus. With a language to describe this relationship,
it is now possible to study quantitatively the effects of genetics and
chemicals on this sensorimotor behavior. We believe that the utility
of the model is quite general and could be applied to different model
system. However, we particularly hope that this work helps further
establish {\em C.\ elegans} as a model for nociceptive research.

\section*{Materials and Methods}

\subsection*{Worm preparation and experiment design}

All worms were grown and maintained under standard conditions
\cite{Brenner1974wn}, incubated with food at 20 $^\circ$C. Well fed worms were
washed twice then gently spun down for 1 minute and the supernatant
discarded by aspiration. For the drug application, 100 $\mu$L of
ibuprofen in M9 at 100 $\mu$M was added to the eppendorf tube. For
the wild-type and mutant data set, M9 was used instead of the drug
solution. Worms were then placed in an incubator for 30 minutes at
20$^\circ$C. After that worms were poured onto a seeded agar plate and
transferred to agar assay plates by a platinum wire pick. These assay
plates were incubated at 20$^\circ$C for 30 minutes, and then the experimental
trials were done within the next 30 minutes. In total $N=201$ worms
for the control group, $N=441$ worms for the ibuprofen group, and
$N=100$ worms for the mutant group (ocr-2(ak47) osm-9(ky10) IV;
ocr-1(ak46)) group were tested. The mutant strain was obtained from
the Caenorhabditis Genetics Center.

The thermal nociceptive stimulation instrument has been described
previously \cite{Mohammadi:2013ku}. In summary, an infrared laser is
directed to heat the head of a freely crawling worm ($\sim0.5$mm FWHM)
on an agar plate. The laser pulse is generated with a randomly chosen
laser current between 0 to 200 mA, with a duration of 0.1 s. The
heating of the worm is nearly instantaneous, and it is directly
proportional to the current, between 0 and 2 $^\circ$C for the
the current range used in our experiments. Worms were stimulated only once and not reused.
The movements of the worms are imaged using a standard
stereomicroscope with video capture and laser control software written
in LabVIEW. For each nociception stimulus trial, a random worm is selected on
the plate and its motion is sampled at 60 Hz for 15 s, and the laser
is engaged 1~s after the start of the video recording.
 
\subsection*{Data Analysis}

The recorded nociceptive response videos were then processed with
Matlab to calculate the time series of the worm centroid motion as
described previously \cite{Mohammadi:2013ku}. All the worms that were
not stimulated near the head or were not moving forward in the
beginning of the video were discarded. Numerical derivatives of the
centroid times series were then taken and filtered with a custom 500
ms Gaussian filter, which was a one-sided Gaussian at the edges of the
recorded time period, becoming a symmetric Gaussian away from the
edges. This removed the noise due to numerical differentiation and
also averaged out the spurious fluctuations in the forward velocity
due to the imperfect sinusoidal shapes of the moving worm. We verified
that different choices of the filter duration had little effect on the
subsequent analysis pipeline. The direction of the velocity was
determined by projecting the derivative of the centroid time series
onto the head-to-tail vector for each worm, with the positive and
negative velocity values denoting forward / backward motion,
respectively.

The filtered velocity profiles needed to be subsampled
additionally. This was because the statistical model of the data,
Eq.~(\ref{eq:fullmodel}), involved covariance matrices of the active
and paused velocity profiles, $\Sigma_p$ and $\Sigma_a$ (note that
velocity profiles are not temporally translationally invariant due to
the presence of the stimulus, thus the full covariance matrix is
needed, and not a simpler correlation function). To have a full rank
covariance matrix, the number of trials must be larger than the number
of time points. Alternatively, regularization is needed for covariance
calculation. The autocorrelation function for all three worm types
showed a natural correlation time scale of $\gtrsim 0.2$ s, whether
the data was pre-filtered or not. Thus subsampling at frequency $>5$
Hz would not result in data loss. Therefore, instead of an arbitrary
regularization, we chose to subsample the data at 12 Hz, leaving us
with 37 data points to characterize the first 3 s of the worm velocity
trace after the stimulus application, $1\le t\le 4$ s since the start
of the trial. Equation~(\ref{eq:fullmodel}) additionally needs
knowledge of $T$, the number of effectively independent velocity
measurements in the profile. This is obtained by dividing the duration
of the profile by the velocity correlation time. An uncertainty of
such procedure has a minimal effect on the model of the experiment
since it simply changes log likelihoods of models by the same factor,
not changing which model has the maximum likelihood.

We then considered limiting the duration of the velocity profile used
in model building: if velocities at certain time points do not
contribute to identification of $I$, they should be removed to
decrease the number of unknowns in the model that must be determined
from data (values of the templates at different time points). The
first candidate for removal was the period of about 10 frames (0.16 s)
after the laser stimulation since worms do not respond to the stimulus
so quickly. However, removal of this time period had a negligible
effect on the model performance, and we chose to leave it
intact. In contrast, starting from 3.3 s (2.3 s after the stimulus) the
fraction of explainable variance drops to nearly zero
(cf.~Fig.~\ref{fig:var}) since many worms already had turned by this
time and resumed forward motion. Therefore, we eventually settled on
the time in the $1.0\dots3.3$ s range for building the model.

Whenever needed, we estimated variance of our predictions by
bootstrapping the whole analysis pipeline \cite{bootstrap}.  For this,
we created 1000 different datasets by resampling with replacement from
the original control dataset and the mutant / ibuprofen
datasets. Control statistical models (the scaling function $f$ and the
velocity templates) were estimated for each resampled control
dataset. Standard deviations of these models were used as estimates of
error bars in Fig.~\ref{fig:The-normalized-active}. For
Fig.~\ref{fig:Delta_Ipred}, we additionally needed to form the closest
control / treatment worm pairs. These were formed between the {\em
  resampled} data sets for all worm types as well. Standard deviations
of $\Delta I_{\rm type}$ evaluated by such resampling were then
plotted in Fig.~\ref{fig:Delta_Ipred} and used to estimate $Z$
scores. Note that such resampling produces data were control /
treatment paired worms have slightly larger current differences than
in the actual data; this leads to our error bars being {\em
  overestimates}.

Model in Eq.~(\ref{eq:bayes}) requires knowing $P(I)$. In principle,
this is controlled by the experimentalist, and thus should be
known. In our experiments, $P(I)$ was set to be uniform. However, as
described above, some of the worms were discarded in preprocessing,
and this resulted in non-uniformly distributed current samples. To
account for this, we used the empirical $P_{\rm emp}(I)$ in our
analysis instead of $P(I)={\rm const}$. In turn, $P_{\rm emp}(I)$ was
inferred using a well-established algorithm for estimation of
one-dimensional continuous probability distributions from data
\cite{Nemenman:2002gl}.

\subsection*{Calculating the template velocities, the covariances, and
  the scaling function }

The template for the paused state $\bu_{p}$ is calculated by taking
the average of all paused velocity profiles for each of the three worm
datasets.  The covariance $\Sigma_{p}$ is then the covariance of the
set of the paused velocity profiles. 

For active worms, we start with fixed putative parameter values
$\cI_1$ and $\cI_2$. We then calculate the active template $\bu_{a}$ and
the covariance matrix $\sum_a$ by maximizing the
likelihood in Eq.~(\ref{Pva})
\begin{align}
\frac{\partial \sum_{i}^{N_{{\rm type},a}}\log P(\bv_i|a,I_i)}
{\partial  \bu_{a}}&\propto\sum_{i}^{N_{{\rm type},a}}\left[\bv_i
               f_{\cI_1,\cI_2}(I_i)-\bu_{a}f^2_{\cI_1\,\cI_2}(I_i)\right]=0,
\\
\frac{\partial \sum_{i}^{N_{{\rm type},a}} \log  P(\bv_i|a,I_i)}
{\partial
  \Sigma_a}&\propto \sum_{i}^{N_{{\rm type},a}}\left[\bv_i-\bu_{a}f_{\cI_1,\cI_2}(I_i)\right]^2-(\Sigma_a)^{-1}=0,
\end{align}
where $N_{{\rm type},a}$ is the number of active worms of the analyzed
type. This gives:
\begin{align}
  \bu_{a}(\cI_1,\cI_{2})&=\frac{\sum_{i=1}^{N_{{\rm
                          type},a}}\bv_if_{\cI_1,\cI_2}(I_i)}{\sum_{i=1}^{N_{{\rm
                          type},a}}f^2_{\cI_1,\cI_{2}}(I_i)},\\
  \Sigma_a &=\sum_{i}^{N_{{\rm type},a}}\left[\bv_i-\bu_{a}f_{\cI_1,\cI_2}(I_i)\right]^2.
\end{align}
Having thus estimated $\bu_a$ and $\Sigma_a$ at fixed parameter values
$\cI_1$, $\cI_2$, we maximize $\prod_iP(\bv_i|a,I_i)$ over the
parameters using standard optimization algorithms provided by
MATLAB. We perform optimization from ten different initial conditions
to increase the possibility that we find a global, rather than the
local maximum.

\subsection*{Acknowledgements}

We thank the granting agencies NSF, HFSP (Human Frontier Science
Program), and NSERC (Natural Sciences and Engineering Council of
Canada) for funding, and the Aspen Center for Physics for hosting the
program on ``Physics of Behavior'', where some of this work was done.


\end{document}